\documentclass{PoS}
\usepackage{multirow}
\usepackage{caption}
\usepackage{url}

\title{Monte Carlo study of a single SST-1M prototype for the Cherenkov Telescope Array}

\ShortTitle{MC study of single SST-1M prototype for CTA}

\newcommand{\krakow}{Krak\'ow{}}

\author{\speaker{J. Jury\v{s}ek}$^{1,2}$, C.~Alispach$^3$, I.~Al~Samarai$^3$, 
M.~Balbo$^4$, A.~Barbano$^3$, V.~Beshley$^{13}$, A.~Biland$^5$, J.~Blazek$^1$, J.~B{\l}ocki$^6$, J.~Borkowski$^{10}$, T.~Bulik$^7$, F.~Cadoux$^3$,
L.~Chytka$^2$, V.~Coco$^3$, N.~De Angelis$^3$, D.~della Volpe$^3$, Y.~Favre$^3$, T.~Gieras$^6$, M.~Grudzi{\'n}ska$^7$, P.~Hamal$^2$, M.~Heller$^3$,
M.~Hrabovsky$^2$, J.~Kasperek$^{11}$, K.~Koncewicz$^6$, A.~Kotarba$^6$, E.~Lyard$^4$, E.~Mach$^6$, D.~Mandat$^1$, S.~Michal$^2$, J.~Micha{\l}owski$^6$, R.~Moderski$^{10}$, T.~Montaruli$^3$, A.~Nagai$^3$, D.~Neise$^5$, J.~Niemiec$^6$, T.R.S.~Njoh~Ekoume$^3$,
M.~Ostrowski$^8$, M.~Palatka$^1$, P.~Pa{\'s}ko$^9$, H.~Przybilski$^6$, M.~Pech$^1$, B.~Pilszyk$^6$, P.~Rajda$^{11}$, P.~Rozwadowski$^7$,
Y.~Renier$^3$, P.~Schovanek$^1$, K.~Seweryn$^9$, V.~Sliusar$^4$, D.~Smakulska$^6$, D.~Sobczy\'{n}ska$^{12}$, {\L}.~Stawarz$^8$,J.~\'{S}wierblewski$^6$,
P.~\'{S}wierk$^6$, P.~Travnicek$^1$, I.~Troyano Pujadas$^3$, R.~Walter$^4$, M.~Wiecek$^6$, A.~Zagda\'{n}ski$^8$, K.~Zi{\c e}tara$^8$, for the CTA consortium\footnote{for consortium list see PoS(ICRC2019)1177} \\
\llap{$^1$}\textit{FZU - Institute of Physics of the Czech Academy of Sciences, 17. listopadu 50, Olomouc \& Na Slovance 2, Prague, Czech Republic.}\\
\llap{$^2$}\textit{Palacky University Olomouc, Faculty of Science, RCPTM, 17. listopadu 50, Olomouc, Czech Republic.}\\
\llap{$^3$}\textit{DPNC - Universit\'e de Gen\`eve, 24 Quai Ernest Ansermet, CH-1211 Gen\`eve, Switzerland}\\
\llap{$^4$}\textit{D\'epartement d'Astronomie, Universit\'e de Gen\`eve, Chemin d'Ecogia 16, CH-1290 Versoix, Switzerland}\\
\llap{$^5$}\textit{ETH Zurich, Institute for Particle Physics and Astrophysics, Otto-Stern-Weg 5, 8093 Zurich, Switzerland}\\
\llap{$^6$}\textit{Institute of Nuclear Physics Polish Academy of Sciences, PL-31342 Krakow, Poland}\\
\llap{$^7$}\textit{Astronomical Observatory, University of Warsaw, Al. Ujazdowskie 4, 00-478 Warsaw, Poland}\\
\llap{$^8$}\textit{Astronomical Observatory, Jagiellonian University, ul. Orla 171, 30-244 \krakow, Poland}\\
\llap{$^9$}\textit{Centrum Bada{\'n} Kosmicznych Polskiej Akademii Nauk,  18a Bartycka str., 00-716 Warsaw, Poland }\\
\llap{$^{10}$}\textit{Nicolaus Copernicus Astronomical Center, Polish Academy of Sciences,  ul. Bartycka 18, 00-716 Warsaw, Poland }\\
\llap{$^{11}$}\textit{AGH University of Science and Technology, al.Mickiewicza 30,  30-059 \krakow, Poland}\\
\llap{$^{12}$}\textit{Department of Astrophysics, University of {\L}\'od\'z, ul. Pomorska 149/153, 90-236 {\L}\'od\'z, Poland }\\
\llap{$^{13}$}\textit{Pidstryhach Institute for Applied Problems of Mechanics and Mathematics, National Academy of Sciences of Ukraine, 3-b Naukova St., 79060, Lviv, Ukraine} \\
E-mail: \email{jurysek@fzu.cz}}

\abstract{
The SST-1M telescope was developed as a prototype of a Small-Size-Telescope for the Cherenkov Telescope Array observatory and it has been extensively tested in Krakow since 2017. In this contribution we present validation of the Monte Carlo model of the prototype and expected performance in Krakow conditions. We focus on gamma/hadron separation and mono reconstruction of energy and gamma photon arrival direction using Machine learning methods.
}

\FullConference{36th International Cosmic Ray Conference -ICRC2019-\\
		July 24th - August 1st, 2019\\
		Madison, WI, U.S.A.}

\begin{document}

\section{Introduction}

SST-1M was developed as a prototype of a Small-Sized Telescope for the Cherenkov Telescope Array \cite{cta_concept}, designed for observations of the gamma-ray induced atmospheric showers for energies above 3~TeV. The SST-1M design is based on Davies-Cotton concept with a 4-m multi-segment mirror dish composed of 18 hexagonal facets \cite{sst-1m}. The telescope is equipped with an innovative camera which features a fully digital readout and trigger system, called DigiCam \cite{heller_digicam} and adopts silicon photomultipliers (SiPMs) as light sensing technology. Photo-detection plane of the camera consists of 1296 SiPM pixels and the whole system has a large 9-degree diameter field of view (FoV).

The first fully operational SST-1M prototype is located in Krakow and has been extensively tested since 2017. In order to understand the data from the prototype, to estimate its performance in Krakow conditions and to perform a high level analysis as gamma/hadron separation or image reconstruction, a precise Monte Carlo (MC) simulation of the telescope is neccessary.

In the following sections a brief description of MC model validation, which is also mandatory for the MC validation process of CTA, is presented (the full description can be found in \cite{simulation_paper}) together with estimation of the prototype performance in mono regime and Krakow atmospheric conditions.

\section{Monte-Carlo model of the SST-1M prototype} \label{sec.mc}

\begin{figure}[!t]
\centering
\begin{tabular}{cc}
\includegraphics[width=.49\textwidth]{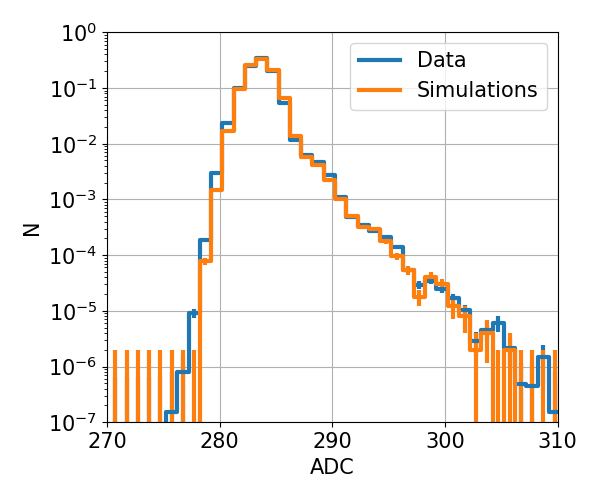} & \includegraphics[width=.49\textwidth]{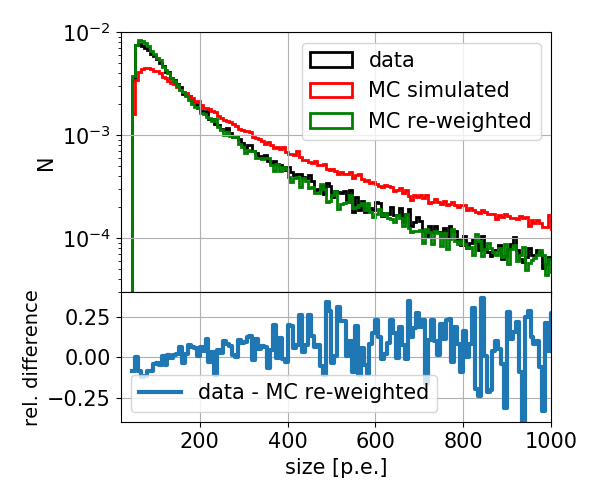}  \\
\end{tabular}
\caption{\textit{Left:} Distributions of raw ADC counts for dark run data and for simulated pedestal events in one selected pixel. \textit{Right:} Distribution of Hillas parameter \textit{size} for data after quality cuts (black), for simulated diffuse protons (red) and for simulations after re-weighting wrt. the real CR spectrum (green).}
\label{fig.raw_adc}
\end{figure}

The MC model of the prototype was created based on the data taken at IFJ in Krakow and the laboratory measurements of individual components like mirrors, window or camera electronics. The gain and electronic noise of individual SiPMs were also measured in the lab, but to check validity of the MC model, raw ADC distributions were compared for a dark run with camera lid closed with \texttt{sim$\_$telarray} \cite{sim_telarray} simulations of pedestal events for a wide parameter space of gain, noise and dark count rate (DCR). The best matching distribution for a single camera pixel is shown in the left Fig.~\ref{fig.raw_adc}. It turned out that the parameters which need to be set in \texttt{sim$\_$telarray} to reproduce the data are slightly different from the laboratory measurements. While the difference in DCR can be explained by temperature differences between the lab and the site, the difference in amplitude and electronic noise ($15 \%$ and $8 \%$ respectively) is still a matter of investigation \cite{simulation_paper}.

\begin{figure}[!t]
\centering
\begin{tabular}{cc}
\includegraphics[width=.49\textwidth]{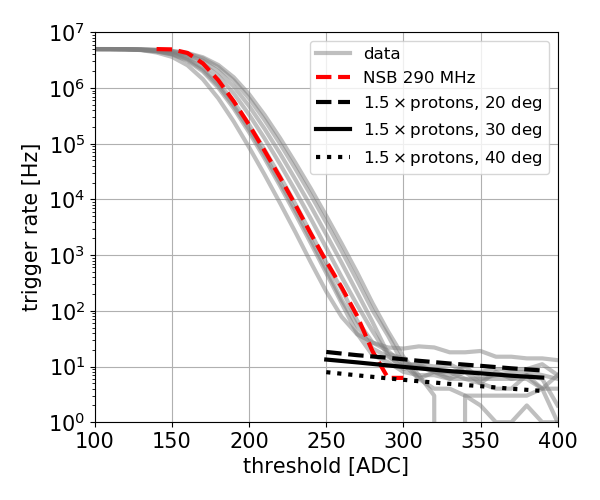} & \includegraphics[width=.49\textwidth]{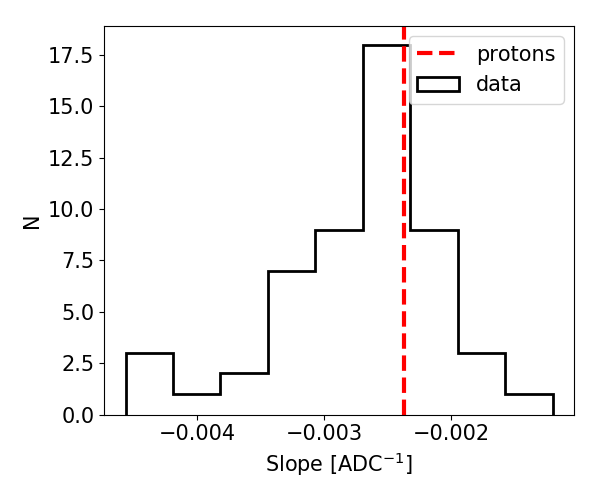}  \\
\end{tabular}
\caption{\textit{Left:} Rate scans taken at different zenith angles (grey solid lines) compared with simulated rate scan for NSB = 290~MHz (red dashed line) and with $1.5 \times$ simulated proton trigger rates for three zenith angles (black lines). \textit{Right:} Distribution of slopes of a linear fit of $\log_{10}{(\mathrm{trigger \;\, rate})} = f(\mathrm{threshold})$ dependency for CR trigger dominated part of the rate scans.}
\label{fig.rate_scans}
\end{figure}

As a high level test of MC model validity, the measured rate scans were compared with simulations, which is shown in the left Fig.~\ref{fig.rate_scans}. The shape of simulated rate scan corresponds well with the data in Night Sky Background (NSB) dominated part of the dependency. In the left Fig.~\ref{fig.rate_scans} there is also shown the proton trigger rate for three zenith angles multiplied by 1.5 as a correction for cosmic rays (CR) composition. Detailed analysis showed that the real trigger rate in that part of the rate scan tends to be about $2\times$ higher than in simulations. The right plot of Fig.~\ref{fig.rate_scans}, however, shows that the slope of a linear fit of $\log_{10}{(\mathrm{trigger \;\, rate})} = f(\mathrm{threshold})$ dependency for CR trigger dominated part of the rate scans is consistent with the slope for simulated protons.

The MC model of the prototype was used to build a small library of \texttt{CORSIKA} \cite{corsika} and \texttt{sim$\_$telarray} simulations of the prototype for Krakow atmospheric conditions for 20 deg zenith angle. The trigger threshold was set on 260~ADC (45.6~p.e.) for the simulations to get about 300~Hz trigger rate for typical NSB level in Krakow (about 300~MHz\footnote{Compare with a typical NSB level at CTA-S site of about 40 MHz.}), which corresponds with the adopted strategy of threshold settings during data taking.

For calibration, cleaning and Hillas parameters \cite{hillas_parameters} extraction, the SST-1M pipeline  \texttt{digicampipe}\footnote{\url{https://github.com/cta-sst-1m/digicampipe}}, which is based on \texttt{ctapipe}\footnote{\url{https://github.com/cta-observatory/ctapipe}}, was used. As an image cleaning method, standard tailcut cleaning is adopted in \texttt{digicampipe}, using two cuts optimized by minimizing the \textit{miss} Hillas parameter on point source gamma simulations.

After processing data from scientific runs and proton simulations, another important high level test of MC validity can be done by comparing the Hillas parameter distributions. One night of ON source observation\footnote{OFF data wasn't taken, but ON data are still dominated by CR as only quality cuts were applied in this case.} was chosen for this test and the distribution of the Hillas parameter \textit{size} is shown in the right Fig.~\ref{fig.raw_adc}. The protons were simulated with -2.0 spectral index and therefore the distribution of parameters had to be re-weighted with respect to the expected trigger rates for the real CR spectrum (see Sec.~\ref{sec.trigger_rates}). After re-weighting, both distributions show very good match.

\section{Mono reconstruction methods and the prototype performance}

\subsection{Arrival direction reconstruction} \label{sec.disp}

\begin{figure}[!t]
\centering
\begin{tabular}{cc}
\includegraphics[width=.49\textwidth]{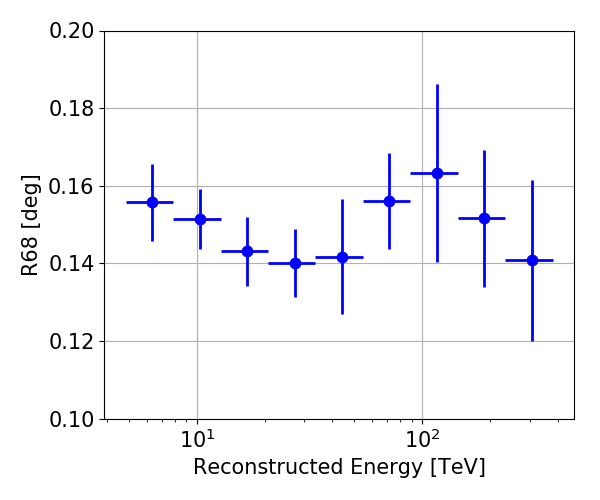} & \includegraphics[width=.49\textwidth]{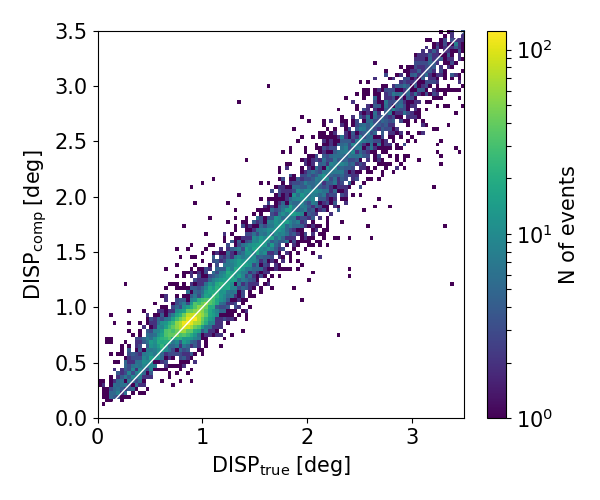}  \\
\end{tabular}
\caption{\textit{Left:} Angular resolution of the prototype in mono regime as a function of reconstructed energy. \textit{Right:} Reconstructed DISP as a function of true DISP.}
\label{fig.angular_resolution}
\end{figure}

For the arrival direction reconstruction of the showers, several implementations of DISP method have been tested. The \textit{DISP} parameter, which is the angular distance between predicted source position and the cleaned image center of gravity, depends mostly on the \textit{length} and \textit{width} Hillas parameters. First, we tested several formulas for \textit{DISP} \cite{disp_lessard, disp_kranich_stark, disp_luke_riley} and minimized parameters in the equations with respect to squared distance of reconstructed source position from the center of the FOV ($\theta^2$) using simulations of on-axis point source gammas. It turned out, that the resolution ($R_{68}$) we can reach using this approach varies from 0.45~deg to 0.60~deg, depending on the formula used. Another approach we tested consist in filling multidimensional look-up tables \textit{DISP} = f[\textit{length},\textit{width}, \textit{size}, \textit{time$\_$gradient}\footnote{A slope of photon arrival times in each pixel along main axis of Hillas elipse.}], which leads to $R_{68}$ = 0.16~deg. Finally, we tested K-Neighbors (KN) and Random Forest (RF) regressors from \texttt{scikit-learn} Python library \cite{scikit_learn} for \textit{DISP} determination and RF classifier to determine the correct sign of \textit{DISP}. The simulation data set was split into train, validation and test dataset and optimal parameters for each method were grid-searched and cross validated. Both methods tested perform similarly according to coefficient of determination (KN: $R^2_\mathrm{test} = 0.945$, RF: $R^2_\mathrm{test} = 0.956$) and the image resolution after Random Forrest regression, which was finally adopted as a primary method of arrival direction reconstruction of the prototype data, is $R_{68}$ = 0.15~deg. The angular resolution as a function of reconstructed energy is shown in Fig.~\ref{fig.angular_resolution}.

\subsection{Energy reconstruction}

The number of emitted Cherenkov photons is proportional to energy of primary gamma ray. Therefore, the energy of primary photon can be reconstructed from the number of photons (or photoelectrons) detected, which is given by the Hillas parameter \textit{size}, and distance of shower core from the telescope, which is expressed by the \textit{impact parameter}. In case of multi-telescope detection, \textit{impact parameter} can be calculated from trivial geometry, which is not possible for mono reconstruction. In that case, however, the \textit{impact parameter} for distant showers can be derived from the \textit{time$\_$gradient} of a shower image.

\begin{figure}[!t]
\centering
\begin{tabular}{cc}
\includegraphics[width=.49\textwidth]{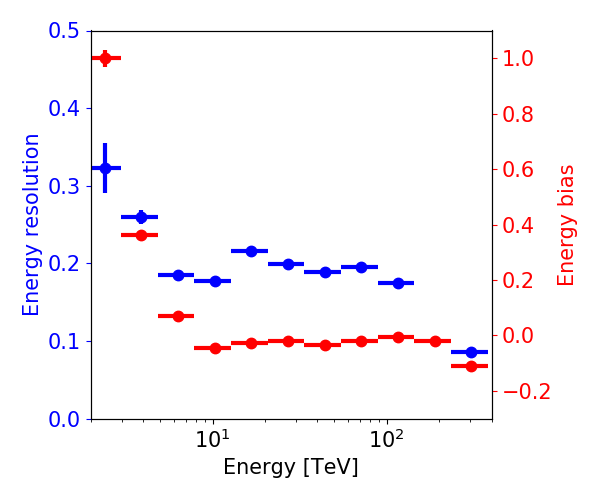} & \includegraphics[width=.49\textwidth]{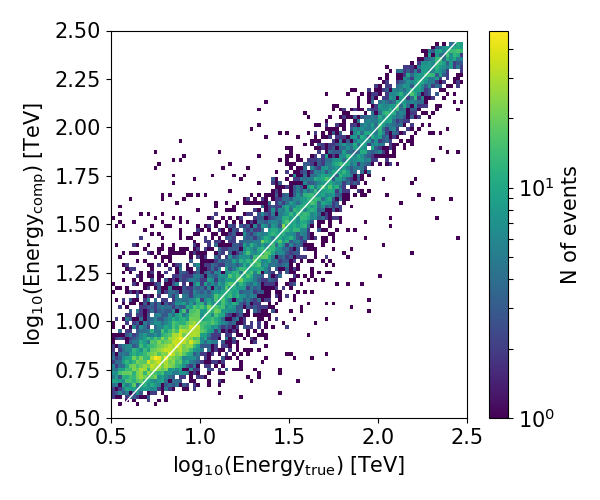}  \\
\end{tabular}
\caption{\textit{Left:} Energy resolution and bias. \textit{Right:} Reconstructed energy as a function of true energy.}
\label{fig.energy_resolution}
\end{figure}

For the energy reconstruction, we tested KN and RF regressors from \texttt{scikit-learn} library, trained on simulated point source gammas. Optimal parameters for each method were found by grid-searching and cross validated. The RF regressor performs slightly better ($R^2_\mathrm{test}=0.943$) than KN regressor ($R^2_\mathrm{test}=0.925$) and it was therefore used as a primary method for energy reconstruction in this study. The correlation between simulated and reconstructed energy is shown in right Fig.~\ref{fig.energy_resolution}

The energy resolution and the bias were calculated from a gaussian fit of the relative error distribution in each true energy bin (see left Fig.~\ref{fig.energy_resolution}). One can see that energy resolution which can be reached in our mono reconstruction is approximately $20 \, \%$ above 10~TeV with $-3\%$ energy bias between 10 and 100~TeV.

\subsection{Gamma-hadron separation}

\begin{figure}[!t]
\centering
\begin{tabular}{cc}
\includegraphics[width=.49\textwidth]{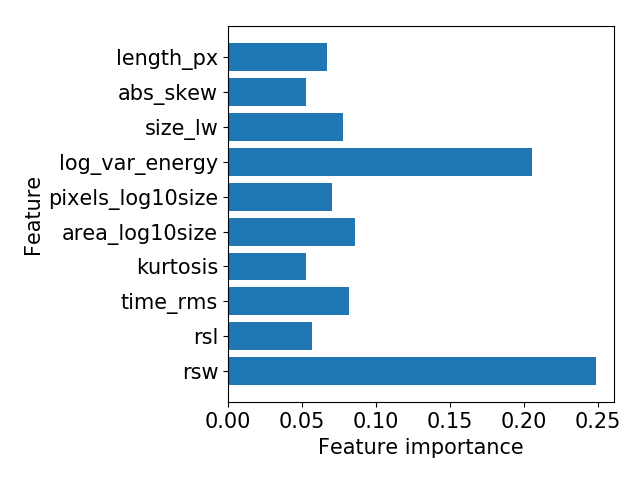} & \includegraphics[width=.49\textwidth]{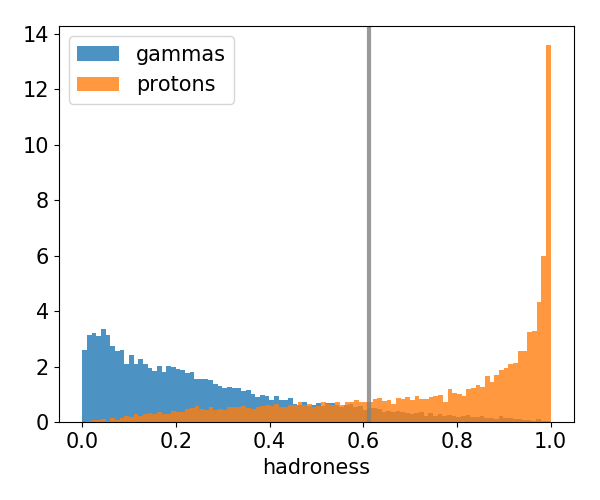} \\
\end{tabular}
\caption{\textit{Left:} Importance of features used. \textit{Right:} Distribution of hadroness for gammas and protons.}
\label{fig.gh_separation_rf_point_features}
\end{figure}

In IACT observations, the trigger rate from diffuse CR background is about $1000\times$ higher than the trigger rate from gamma ray photons and therefore a strong background suppression is necessary. In this analysis, we used RF classifier from \texttt{scikit-learn}, trained on diffuse protons and point source gammas after quality cuts\footnote{\textit{width} / \textit{length} > $10^{-3}$, \textit{n$\_$islands} = 1, \textit{n$\_$pixels} > 2, \textit{size} > 50~p.e., \textit{leakage2} < 0.3}. Optimal parameters were grid-searched using area under ROC curve (AUC) as a measure of classification performance. The optimal  cut on hadronness was found by maximizing the F1-score \cite{f-score}. In the left Fig.~\ref{fig.gh_separation_rf_point_features} one can see that the most important features for classification are Reduced Scaled Width and $\log(\mathsf{var}(E))$, which is the inter-tree variance of reconstructed energy, higher in general for protons than for gammas as a consequence of the fact that RF for energy reconstruction was trained on gamma events only \cite{temme}. The optimal performance of the separation gives \textit{AUC} = 0.905, \textit{precision} = 0.762 and \textit{recall} = 0.908. Having false positive rate of $28\%$ the separation is weaker than we expected, but our preliminary tests with the use of more powerful boosted decision trees methods lead to similar performance, which suggests that such performance is close to mono reconstruction limits of the telescope.

\subsection{Expected trigger rates and differential sensitivity}  \label{sec.trigger_rates}

\begin{figure}[!t]
\centering
\begin{tabular}{cc}
\includegraphics[width=.49\textwidth]{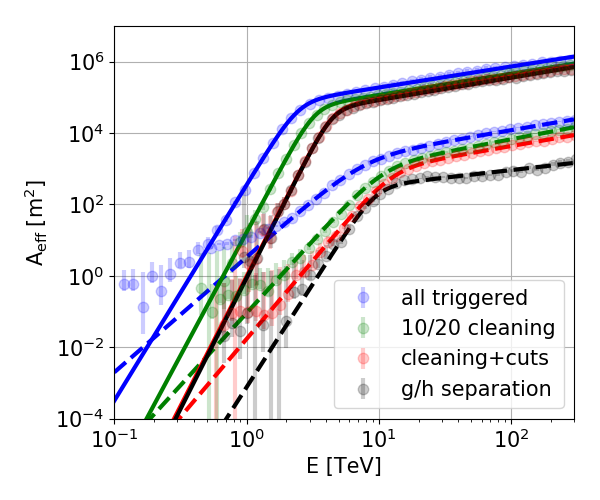} & \includegraphics[width=.49\textwidth]{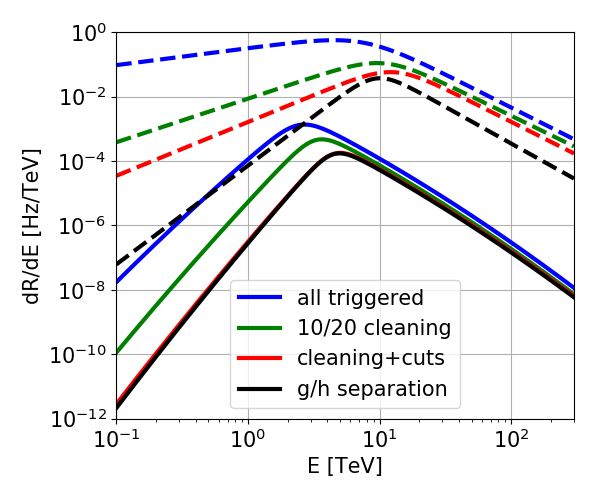} \\
\end{tabular}
\caption{Effective areas (\textit{left}) and expected event rates (\textit{right}) for Crab (full lines) and diffuse protons (dashed lines) at 20~deg zenith angle shown for all analysis steps.}
\label{fig.ea_event_rates}
\end{figure}

In order to estimate the sensitivity of the prototype for Krakow conditions with high NSB level, the effective areas and the expected event rates for a gamma point source with Crab spectrum and background diffuse protons with CR spectrum were calculated. In Fig.~\ref{fig.ea_event_rates}, effective areas and differential event rates are plotted for each analysis step: all triggered events, events that survived cleaning, events after cleaning and quality cuts and finally, events after gamma/hadron separation. The energy thresholds for gamma ray showers detection and the integrated event rates for the full telescope FoV are listed in Tab.~\ref{tab.event_rates}. The sensitivity, calculated with respect to three criteria on the flux ($5\sigma$ detection $\&$ $N_\gamma > 5\%$ of the residual background $\&$ $N_\mathrm{excess} > 10$) with an additional cut on distance from the source of 0.15 deg, shows that even in Krakow a solid $5\sigma$ detection of Crab should be possible for 50h observation.

\begin{table}
	\begin{minipage}{.49\linewidth}
		\centering
		\begin{tabular}{cccc} 
		\hline 
		 & Energy 	& Crab  & Proton \\
		 & threshold & event rate  & event rate \\
		 & [TeV] & [mHz] &  [Hz] \\
		\hline 
		All & \multirow{2}{*}{2.692} & \multirow{2}{*}{5.722} & \multirow{2}{*}{8.957} \\
		triggered & & & \\
		\hline 
		After & \multirow{2}{*}{3.641} & \multirow{2}{*}{2.447} &  \multirow{2}{*}{2.678} \\
		cleaning & & & \\
		\hline 
		Quality & \multirow{2}{*}{5.009} & \multirow{2}{*}{1.237} & \multirow{2}{*}{1.558} \\
		cuts & & & \\
		\hline 
		g/h & \multirow{2}{*}{4.966} & \multirow{2}{*}{1.174} & \multirow{2}{*}{0.605} \\
		separation  & & & \\
		\hline 
		\end{tabular}
		\caption{Expected event rates for the full telescope FoV and energy thresholds for Crab observation in Krakow.}
		\label{tab.event_rates}
	\end{minipage}\hfill
	\begin{minipage}{.49\linewidth}
		\centering
		\includegraphics[width=1.0\textwidth]{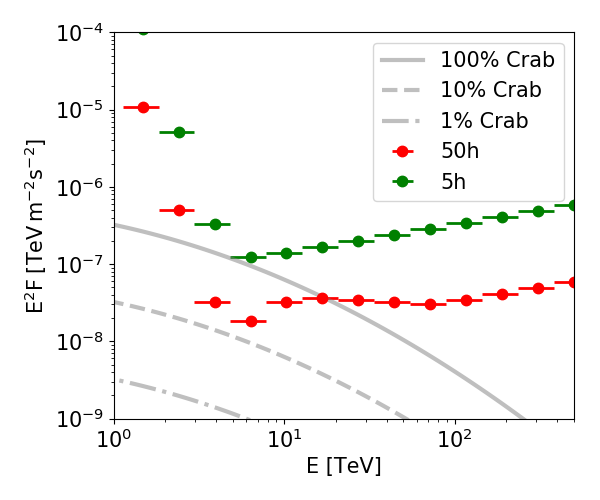}
		\captionof{figure}{Differential sensitivity of SST-1M prototype in Krakow in mono regime.}
		\label{fig.sensitivity}
	\end{minipage}
\end{table}

\section{Acknowledgments}

We gratefully acknowledge financial support from the agencies and organizations listed here: http://www.cta-observatory.org/consortium$\_$acknowledgments. The work is supported by the projects of Ministry of Education, Youth and Sports: MEYS LM2015046, LTT17006 and EU/MEYS CZ.02.1.01/0.0/0.0/16\_013/0001403, Czech Republic, and by the grant Nr. DIR/WK/2017/12 from the Polish Ministry of Science and Higher Education. We greatly acknowledge financial support form the Swiss State Secretariat for Education Research and Innovation SERI.

\end{document}